\documentstyle[aps,twocolumn,epsfig,floats]{revtex}

\begin{document}

\hfuzz=4pt

\twocolumn[\hsize\textwidth\columnwidth\hsize\csname @twocolumnfalse\endcsname

\hbox to \hsize{
\font\fortssbx=cmssbx10 scaled \magstep1
\hbox{\fortssbx University of Wisconsin - Madison}
\hfill$\vcenter{\tighten
                 \hbox{\bf MADPH-99-1097}
                 \hbox{\bf hep-ph/9901387}
                \hbox{January 1999}}$}

\vspace*{2mm}

\title{On Forward $J/\psi$ Production at Fermilab Tevatron}

\author{O.\ J.\ P.\ \'Eboli$^{1,2}$,
E.\ M.\ Gregores$^1$, and F.\ Halzen$^1$}

\address{\vspace*{2mm}
$^1$Department of Physics, University of Wisconsin,
Madison, WI 53706, USA\\[1mm]
$^2$Instituto de F\'{\i}sica Te\'orica,
Universidade Estadual Paulista,
Rua Pamplona 145, S\~ao Paulo, SP 01405-900, Brazil}

\maketitle

\vspace*{-2mm}

\begin{abstract}

The D$\O$ Collaboration has recently reported the measurement of $J/\psi$
production at low angle. We show here that the inclusion of color octet
contributions in any framework is able to reproduce this~data.

\end{abstract}

\bigskip]
\narrowtext


The D$\O$ collaboration has recently reported the first measurement of
$J/\psi$ production in the forward pseudorapidity region $2.5 \leq |\eta| \leq
3.7$ in $p\bar{p}$ collisions at $\sqrt{s}=1800$ GeV \cite{d098}. It was shown
that the dependence of the $J/\psi$ cross section with its transverse momentum
confirmed theoretical expectations based on NRQCD \cite{nrqcd}. Here we show
that the soft color model \cite{us-1} is also able to explain these
results. The implication of this result is that this data, once more, requires
the inclusion of color octet perturbative diagrams for the production of
$\psi$'s.  How this is implemented is not decisive
\cite{us-1,us-2}.

\begin{figure}
\begin{center}
\mbox{\epsfig{file=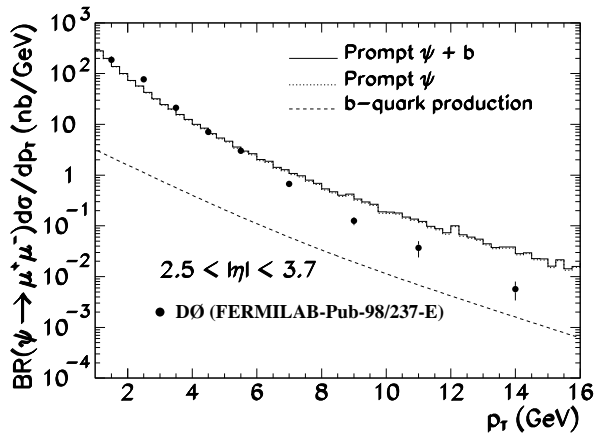,width=0.9\linewidth}} 
\end{center}
\caption{The $p_T$ dependence of the differential cross section} 
\label{fig:pt} 
\end{figure}

We have evaluated $\psi$ production following reference
\cite{us-1}. Like the measurement, we included prompt production, as well
as production via $b$-decay. We only adjusted the renormalization and
factorization scales as appropriate for a leading order calculation.  The
predictions of the soft color model for the forward $J/\psi$ production at the
Tevatron is compared with the experimental results in Fig.~\ref{fig:pt}. As
can be seen, the leading order evaluation of the soft color model adequately
describes the shape of the forward $p_T$ distribution and its absolute
normalization.

\begin{figure} 
\begin{center}
\mbox{\epsfig{file=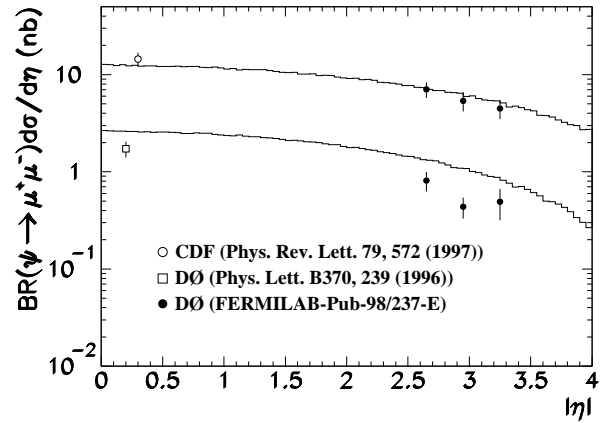,width=0.9\linewidth}} 
\end{center}
\caption{The pseudorapidity dependence of the differential cross
section} 
\label{fig:eta} 
\end{figure}

It is interesting to verify that the soft color model describes the
rapidity distribution for different cuts on $p_T$. The result is shown
in Fig.~\ref{fig:eta}, and a good agreement is obtained. 

\acknowledgments
\unskip\smallskip  

This research was supported in part by the University of Wisconsin Research
Committee with funds granted by the Wisconsin Alumni Research Foundation, by
the U.S.\ Department of Energy under grant DE-FG02-95ER40896, by
Funda\c{c}\~{a}o de Amparo \`a Pesquisa do Estado de S\~ao Paulo (FAPESP), and
by Conselho Nacional de Desenvolvimento Cient\'{\i}fico e Tecnol\'ogico
(CNPq).



\end{document}